\documentclass[prl,twocolumn,aps,10pt,showpacs]{revtex4-1}
\usepackage{hyperref,amsmath,graphicx}
\begin{document}
\title{  Low temperature dissipation scenarios in palladium nano-mechanical resonators}
\author{S. Rebari} \author{Shelender Kumar} \author{S. Indrajeet}\altaffiliation{present address: Department of Physics, Syracuse University, Syracuse, NY 13244, USA} \author{Abhishek Kumar}\altaffiliation{present address: NEST, Istituto Nanoscienze-CNR and Scuola Normale Superiore, Piazza San Silvestro 12, 56127 Pisa, Italy} \author{Satyendra P. Pal}\altaffiliation{present address: Department of Physics, Indian Institute of Technology, Hauz khas, New Delhi 110 016 India } \author{A. Venkatesan}\email[e-mail: ]{ananthv@iisermohali.ac.in }
\affiliation{Department of Physical Sciences IISER Mohali, Knowledge city,Sector 81, SAS Nagar, Manauli P.O. 140306, India }
\author{D. Weiss}
\affiliation{Experimental and Applied Physics, University of Regensburg, D-93040 Regensburg, Germany}
\begin{abstract}
We study dissipation in Pd nano-mechanical resonators at low temperatures in the linear response regime. Metallic resonators have shown characteristic features of dissipation due to tunneling two level systems (TLS). This system offers a unique tunability of  the dissipation scenario by adsorbing hydrogen ($H_2$) which induces a compressive stress. The intrinsic stress is expected to alter TLS behaviour. We find a sub-linear power law $\sim T^{0.4}$ in dissipation. As seen in TLS dissipation scenarios we find a logarithmic increase of frequency characteristic from the lowest temperatures till a characteristic temperature $T_{co}$ is reached. In samples without $H_2$ the $T_{co} \sim 1K$ whereas with $H_2$ it is clearly reduced to $\sim 700 mK$. Based on standard TLS phenomena we attribute this to enhanced phonon-TLS coupling in samples with compressive strain. We also find with $H_2$ there is a saturation in low temperature dissipation which may possibly be due to super-radiant interaction between TLS and phonons. We discuss the data in the scope of TLS phenomena and similar data for other systems. 

\end{abstract}

\maketitle

Nano-electromechanical systems (NEMS) are not only sensitive transducers but also form an excellent platform to explore  basic physical phenomena. The spectrum of phenomena include potential macroscopic quantum states\cite{schwab}, electron-phonon coupling\cite{weig} and  mechano-spintronic phenomena\cite{spin_flip}. Scaling of NEMS to smaller sizes (or higher frequencies) results in higher dissipation\cite{ekinci}. Typically GHz frequency devices that can satisfy the rudimentary quantum condition $\hbar \omega \geq k_{B} T$ at dilution fridge temperatures are limited by intrinsic losses despite geometric aspects like clamping loss at boundaries are overcome by free-free resonators designs\cite{huang} or small phonon bottle neck devices\cite{gaid}. 
Photon pressure in microwave cavities can also squeeze MHz frequency NEMS  resulting in occupation numbers close to the ground state of a quantized harmonic oscillator\cite{lehnert}. So far only one dilation-mode resonator has shown  evidence for macroscopic quantum behaviour in mechanical systems\cite{connel}. Intrinsic loss mechanisms in NEMS are not yet fully understood and are crucial to understand de-coherence of quantum phenomena\cite{remus}. \\
At temperatures  below $4.2\textrm{ K}$ almost all materials freeze except helium which forms a quantum liquid due to overlap of nuclear wave-functions\cite{enss}. One may naively expect most solids will show uninteresting behaviour in mechanical response at these temperatures. On the contrary the mechanical responses of solids do vary vibrantly even at temperatures down to even below $T\leq 3 \textrm{ }mK$\cite{feffer}. Tunneling two level systems (TLS) have been used to model low temperature mechanical dissipation in bulk solids successfully\cite{enss,feffer,esqui,hunk_akr}.
\begin{figure}[hb]
\scalebox{0.35}{\includegraphics{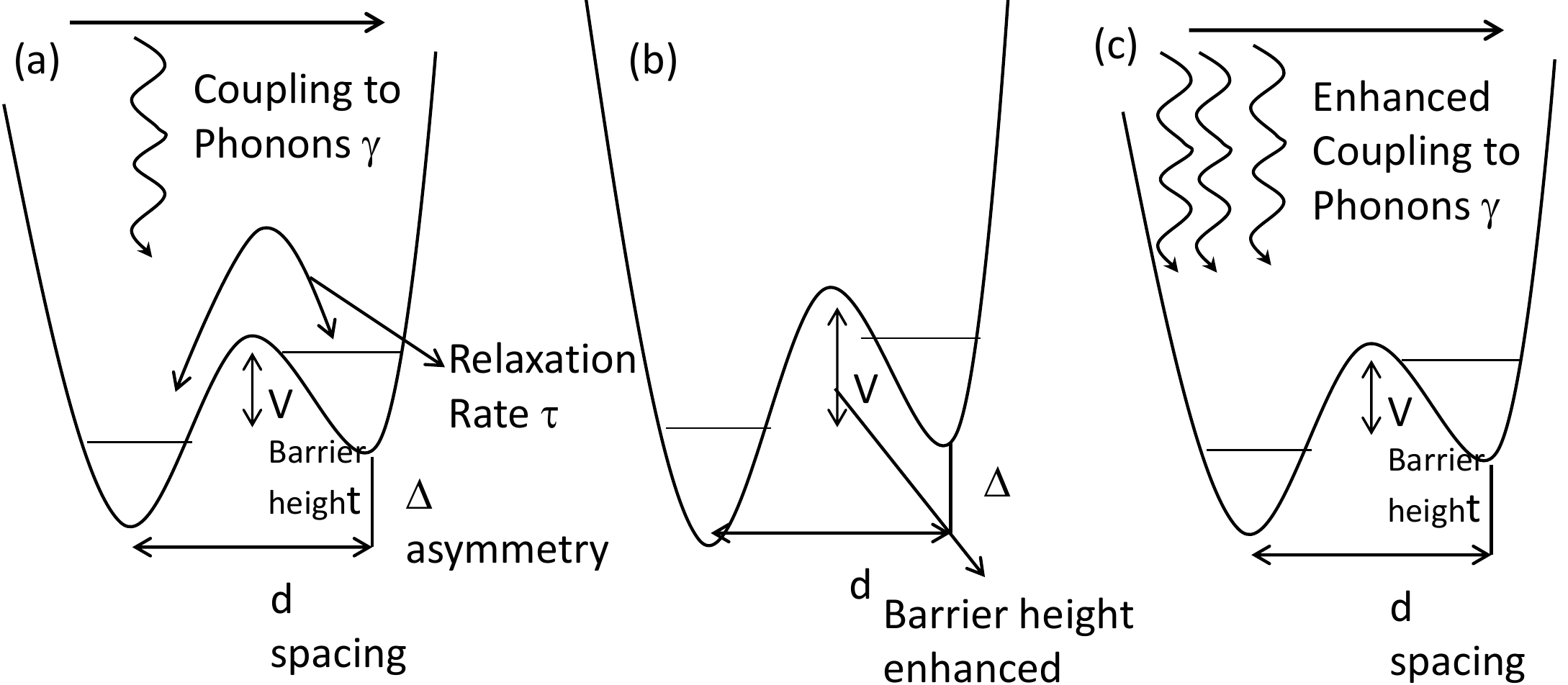}}
\caption{ \small (a) Schematic of a TLS. The key parameters are  barrier height $V$ the barrier asymmetry $\Delta$ the relaxation time $\tau$ and the TLS phonon coupling parameter $\gamma$. The tunnel amplitude $\Delta_{0} \sim \hbar \Omega e^{-\lambda} \textrm{ where } \lambda= d\sqrt{2mV/\hbar^2}\textrm{ } m \textrm{ is particle mass like term.} $ TLS splitting energy $E^{2} = \Delta^{2}+ \Delta_{0}^{2}$. Possible changes in TLS parameters (b) increase of barrier height (c) enhanced phonon coupling.}\label{tls}
\end{figure}
Some of the key parameters of a phenomenological TLS is shown as a schematic in Fig(\ref{tls}). Apart from the parameters for an isolated TLS as shown in Fig(\ref{tls}) the overall energy landscape of TLS like the distribution function for TLS energies $\bar{P}(E,\Delta_{0})$,  distribution of typical relaxation times $\tau_s$ for tunneling (both $\bar{P}(E,\Delta_{0})$ and $\tau_s$ are functionally related)   and how the TLS  interact with each other and interaction with quasi-particles like phonons gives rise to unique behaviour for various classes of systems. These models have been successful in explaining many experimentally accessible properties of bulk solids both amorphous and crystalline. The experimentally studied properties usually have  a Kramers-Kroning type dispersion for a susceptibility $\chi = \chi_{r} + i \chi_{im}$ containing a real (dissipative) and imaginary (dispersive) response, the dissipation $Q^{-1}$ and relative frequency shift $df/f_{0}$ for measurements in this work. 
In systems like amorphous silicon films hydrogenation resulted in lower dissipation possibly due to alteration of the co-ordination of the amorphous network\cite{liuh2}. Recent studies have also shown  density dependent voids to scale with TLS density\cite{hellman}. TLS models for amorphous glasses are general enough to be mapped to several  crystalline and polycrystalline systems. In polycrystalline solids one may map the  variables to several potential TLS candidates like grain boundary angles, kinks and dislocations but with different energy scales for parameters like phonon TLS coupling, density of TLS \textit{etc}. The analogy  is like crystalline materials showing spin-glass states \cite{anderson_varma}. Hence it is not hard to conceive  of pseudo-spin glass states for TLS scenarios. TLS models predict a range of values for quantities like mechanical dissipation \textit{e.g.} $Q^{-1} \sim 10^{-3} -10^{-5}$ for amorphous dielectrics with some exceptions in stressed systems like silicon nitride\cite{parpia1}. The possibility of universality in semiconducting NEMS was suggested in ref\cite{mohanty_rev}. Although TLS models have been successful in explaining behaviour of various bulk solids, there are still open problems like probing TLS at extremely low and high frequencies. What are the physical attributes of TLS? , how to tune them?  are some open questions\cite{leggett,yu}. 
In case of mesoscopic systems the surface to volume ratio and small size of the system complicate modeling them with the well established theories for bulk systems. Hybrid NEMS  on materials like  $Si\textrm{, }GaAs $ and diamond  with metal electrodes for  actuation have been extensively studied with evidence for dissipation due to TLS\cite{mohanty_rev}. A recent work with $Al$ electrodes on $Si$ structures showed a profound difference when measured in the superconducting and non-superconducting state \cite{lulla}. Stand alone metallic nano-mechanical systems are indeed simpler systems to study dissipation. It was demonstrated that tensile stress in these systems increases the quality factor (Q-factor)\cite{li}. In ref\cite{gold} gold nano-mechanical resonators were studied showing evidence for TLS mechanisms. TLS mechanisms possibly due to quasi  1-D phonon mediated dissipation was seen in aluminum in normal\cite{hoehne} and superconducting states\cite{hakonen}. 

In this work we report our studies on mesoscopic $Pd$ beams. The motivation for studying $Pd$ is to probe a system where dissipation scenarios may be modified significantly intrinsically  without external dissipation dilution\cite{tls_stress}.  Palladium's affinity to adsorb $H_2$ is well known. In ref \cite{h2} nano-scale $Au-Pd$ beams have been used as hydrogen sensors by probing frequency shifts due to adsorbed $H_2$ .The $H_2$ not only covers the surface but also forms $H^+$ ions that occupy interstitial sites in the $Pd$ resulting in a compressive strain on the $Pd$ lattice structure. Compressive or tensile stress can affect the barrier height $V$ as in Fig(\ref{tls}.b) or TLS phonon coupling constant $\gamma$ as in Fig(\ref{tls}.c). Metallic beams at cryogenic temperatures have intrinsic tensile stress due to differential thermal contraction with respect to the substrates. Our goal is to tune this tensile stress with exposure to $H_2$ thereby modifying TLS scenarios.  Intrinsic tensile stress is known to drastically alter the TLS scenario in systems like silicon nitride\cite{parpia1,tls_stress}.

The experiments were carried out in a cryo-free dilution fridge. A separate brass vacuum can with home-made RF 
feed-through were used to introduce an exchange gas of $H_2$. Typical samples had a length ($l$) of $4-5$ $\mu m$ , thickness ($t$) around $80$ nm and a width ($w$) of $450-470$ nm were fabricated by e-beam lithography on $Si/SiO_2$ wafers and undercutting the $SiO_2$ in buffered oxide etch. The samples were bonded by mechanically pressing indium coated gold wires on to the chip and to micro-strip tracks with RF launchers.We used a standard magneto-motive technique to probe the resonant response. RF current from a vector network analyzer was driven through the sample with a magnetic field parallel to the wafer plane to excite and detect out of plane motion of the beam due to the Lorentz force\cite{cleland}. 

In initial trials we found samples did not survive thermal cycling to room temperature. Hence a set of two samples Pd4B1L ($\sim4.35\mu m\times390nm$ $l\times w$)  and Pd4B1R ($\sim 4.35\mu m \times 366nm $  $l\times w$) forming  a RF bridge were cooled in $H_2$ exchange gas of $~10^{-3}$ torr and again with $10^{-2}$  torr and subsequently pumped below $10^{-4}$ torr when the mixing chamber temperature was below $160\textrm{ K}$. In the second round of exposure to $H_2$ $Pd4B1R$ was heated with a $0.5 \textrm{ }\mu A$ low frequency current from room temperature down to $160K$. This was unstable in frequency with time-scale several hours to few days possibly due to excessive adsorption of  $H_2$ causing additional diffusion induced dissipation\cite{atalaya,krim} and the data is not discussed here. The frequencies of the samples were estimated as in ref\cite{li} accounting for tension due to differential thermal contraction of the substrate and sample at $4.2 K$. In both cases the resonant frequency was less than the estimated 24 MHz. 

A second set of two samples $Pd2C3L$ ($4.5 \mu m \times 430nm$ $l\times w $ ) and  $Pd2C3R$ ($4.5 \mu m \times 470nm$  $l \times w$) were studied in the absence of $H_2$ separately after pumping the system to $\sim 4\times10^{-5}$ torr over one day and  cooled down while pumped continuously with a turbo pump. The samples had resonant frequencies of $19$ MHz and $29$ MHz.  Although sample dimensions were comparable, $Pd2C3R$ trapped some indium in the etched region below it thereby reducing its effective length to $\sim 3.5 \mu m$  and the predicted frequency matched for this length. An electron microscope image after measurement along with an EDS scan confirmed presence of Indium. The first sample showed some buckling that possibly explains a reduced frequency from the  estimated $23$ MHz. 

A Lorentzian fit to the real and imaginary part of the response was used to extract the loaded Q-factor $Q_l$ and resonant frequency $f_l$. As expected in standard magneto-motive technique the eddy current damping showed a  linear dependence for $B^{2} \textrm{ vs } Q_{l}^{-1} $ in all cases satisfying the relation $Q_{l}^{-1} = Q_{0}^{-1} (1+ \alpha B^{2} )= Q_{0}^{-1} \left[1+ \frac{R_{m} \Re{Z_{ext}}}{\lvert Z_{ext}^{2}\rvert}\right]$ where $R_{m}=\frac{\epsilon l^{2}B^{2}Q_0}{2\pi f_{0} m} $ is the mechanical equivalent of resistance depending on resonator parameters( length, frequency and mass and intrinsic Q-factor $Q_{0}$). The frequency squared also showed a quadratic dependence in field due to presence reactive components like a bias-tee that protected the samples from static charge.  The field dependence was used to estimate the intrinsic $Q_0$ and frequency $f_0$ from loaded values measured at $4$ T. There was not a significant change in the magneto-motive damping parameter $\alpha$ when $H_2$ was added with all samples showing $\alpha$ was $ 1.33-4\times 10^{-6} /T^{2} $ and non monotonic with $H_2$. The optimal power to measure in the linear response regime was $\sim -90 \textrm{ dBm}$ (when pre-cooled with $\sim2\times10^{-3} $ torr  of $H_2$ , hereafter referred as low $H_2$) and $\sim -110$ to $-105 \textrm{ dBm}$ (when pre-cooled with $\sim10^{-2}$ torr $H_2$, hereafter referred to as high $H_2$ ) and $\sim -80 \textrm{ dBm} $ when cooled without $H_2$. In all cases the $H_2$  exchange gas was reduced to a pressure of $\sim 10^{-4}$ torr when the system reached a temperature $\sim 160K$. This change in linear response regime clearly indicates softening of the beams with $H_2$ due to additional compressive stress. 
 
\begin{figure}[b]
\centering
\scalebox{0.4}{\includegraphics{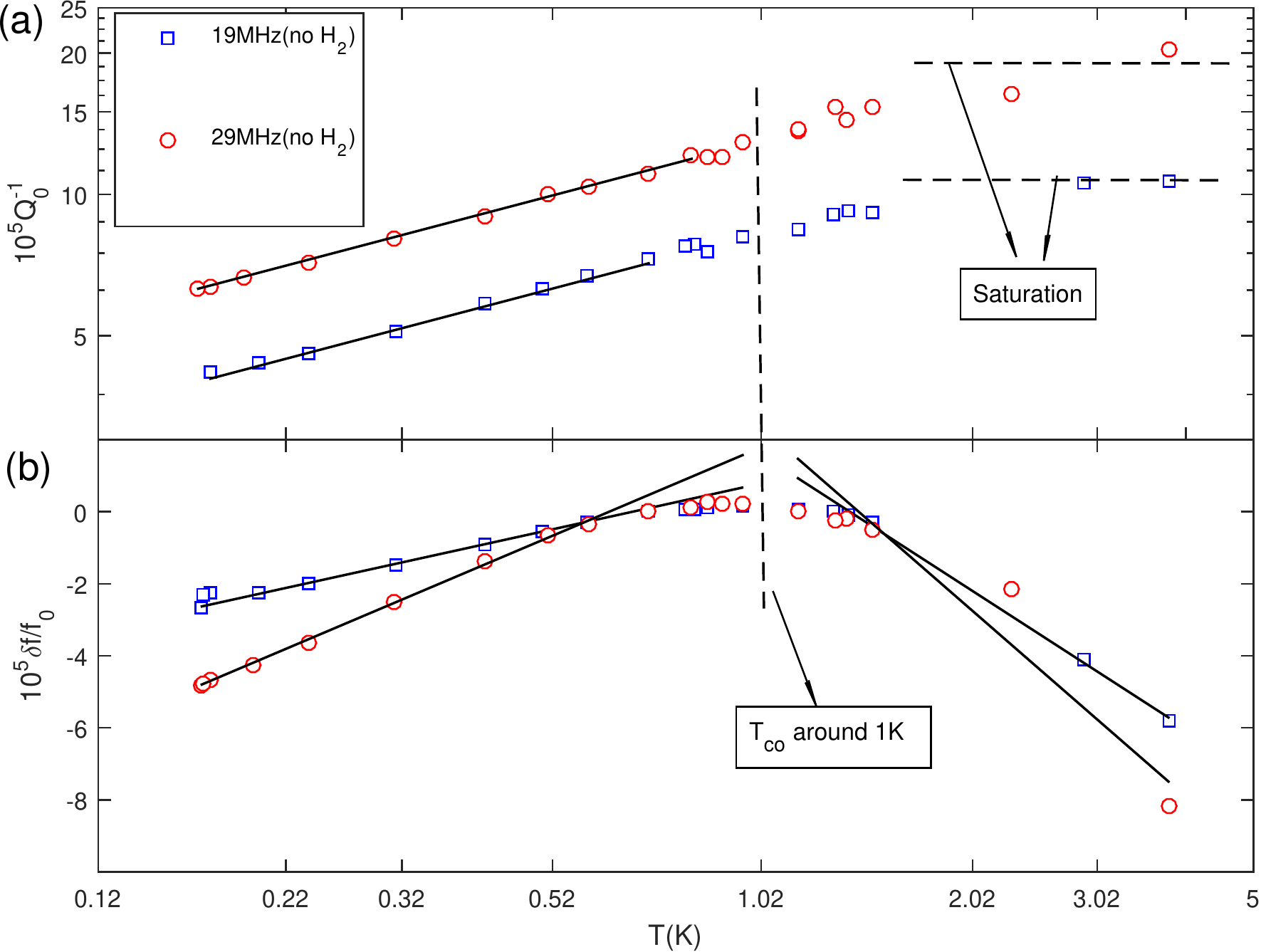}}
\caption{\small (a)Damping $Q^{-1}$ as a function of temperature on the top graph. A power law of $\sim T^{0.39}$ for the $19$ MHz  and $\sim T^{0.41}$ for the $29$ MHz sample is shown as a guide. (b)$\frac{df}{f_0}$ in these samples with a  reference frequency $f_{0}$ at an arbitrary temperature $T_{0}$.  The logarithmic slope quantifying the parameter below $T_{co}\sim 1K $ is $C$ is $\sim 1.85 \times 10^{-5} $  for the $19$ MHz  and $\sim 3.5 \times 10^{-5} $  for the $29$ MHz samples. 
 }\label{a}
\end{figure}

\begin{figure}

\scalebox{0.40}{\includegraphics{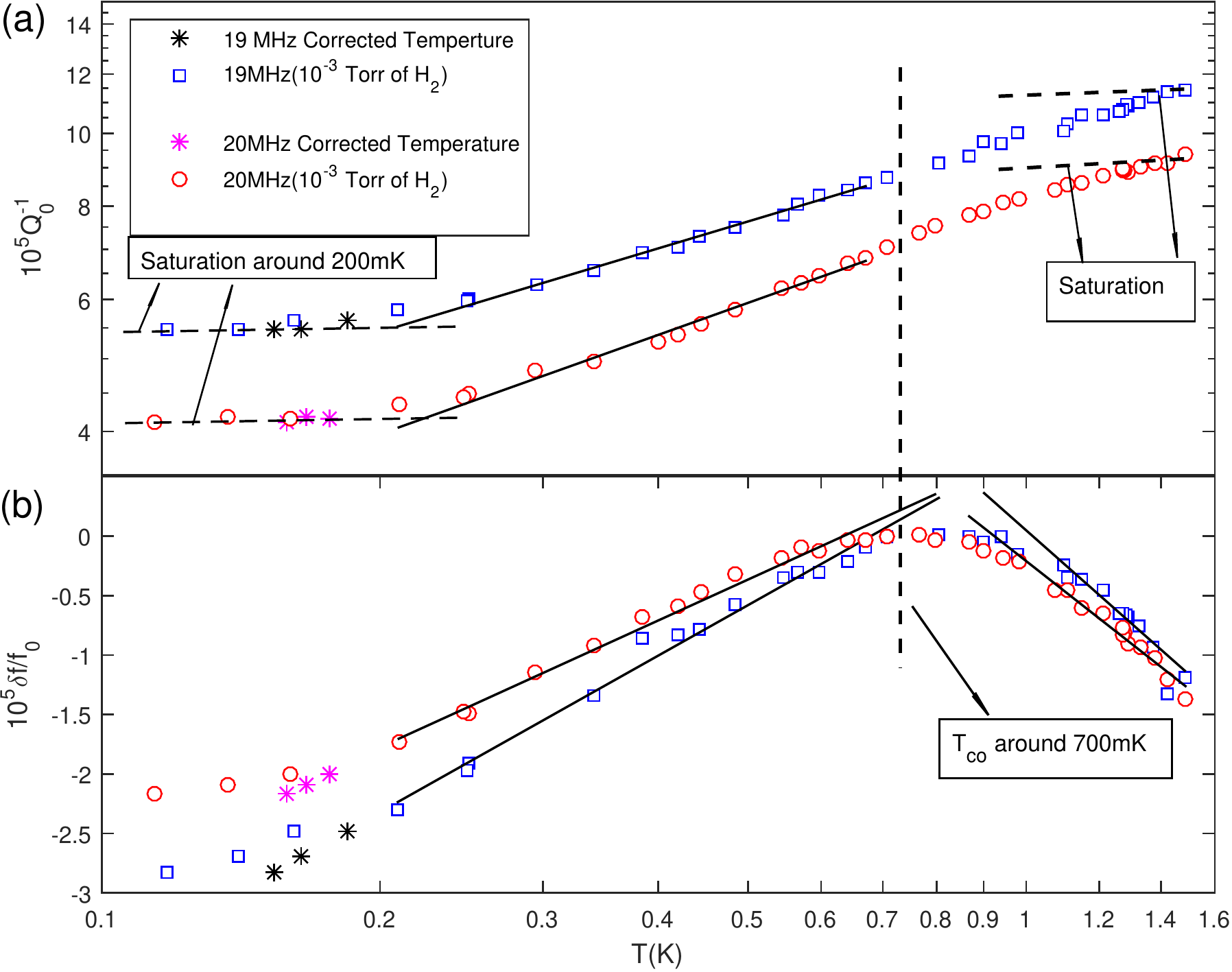}}
\caption{\small (a)$Q^{-1}$ of two samples exposed to $\sim10^{-3}$ torr $H_{2}$ during cool-down. A sub-linear fit with $\sim T^{0.37}$ for the $19$ MHz and $\sim T^{0.43}$ for the $20$ MHz sample is shown for reference. (b) $\frac{df}{f_0}$ for both these samples are shown
with a logarithmic fit of $1.9\times 10^{-5}$ and $1.6\times 10^{-5}$. The last three points show a small deviation from the log fit are corrected by using fit from higher temperatures shown by $*$ symbols. The $Q^{-1} $ in (a) retains the low temperature saturation despite the corrections.    }\label{b}
\end{figure}
\begin{figure}[t]
\scalebox{0.40}{\includegraphics{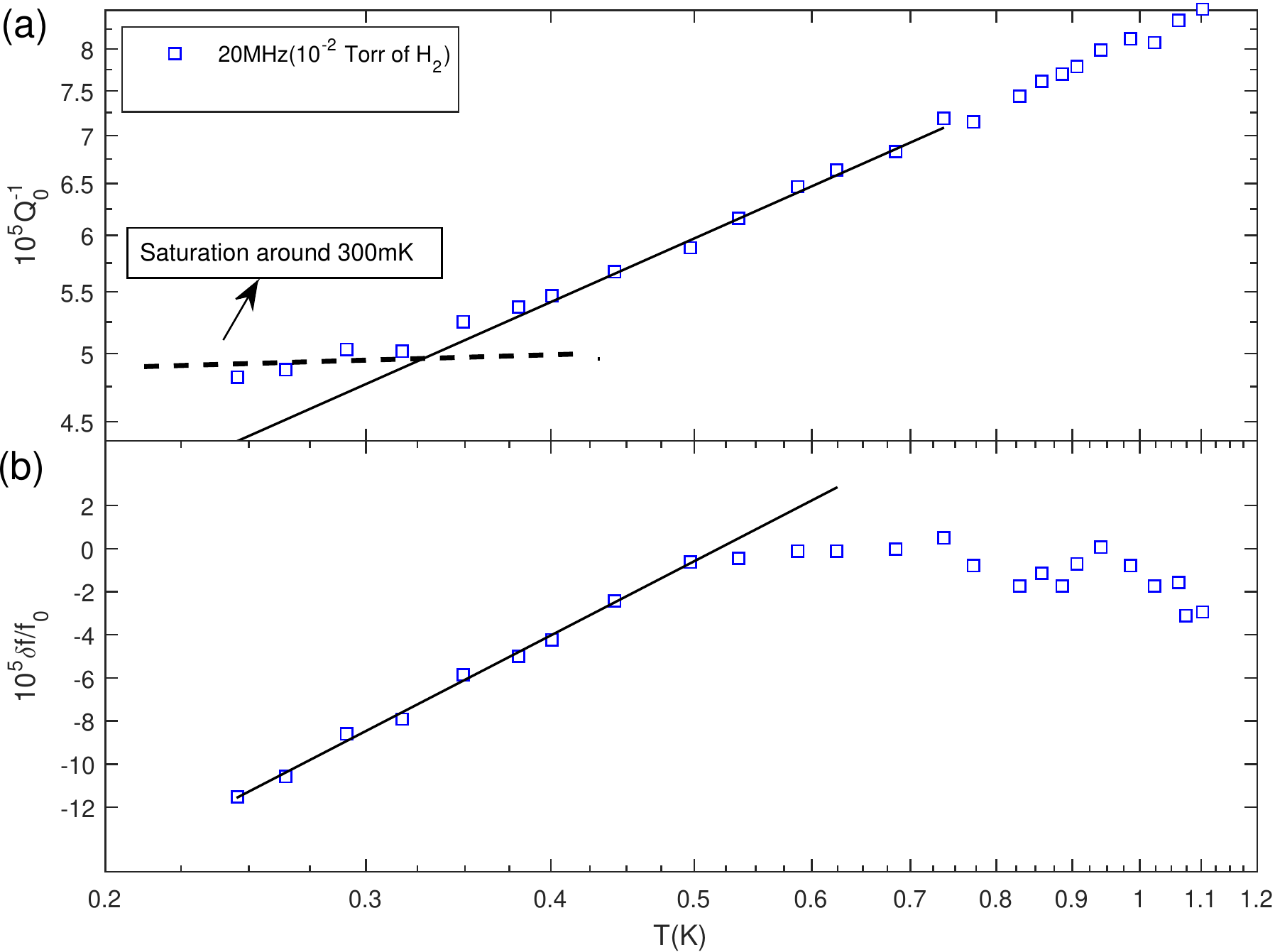}}
\caption{\small (a)The dissipation of the stable $20$ MHz samples exposed to $\sim10^{-2}$ torr $H_2$ during cool-down. A sub-linear fit with $\sim T^{0.43}$ is shown for reference. (b) $\frac{df}{f_{0}}$ with a log fit of $1.54 \times 10^{-5}$ }\label{c} 
\end{figure}

The data in the absence of $H_2$ for the $19$ \& $29$ MHz samples are presented in Fig.\ref{a}. The frequency shift shows a logarithmic increase with raise of temperature till a characteristic temperature $T_{co}\sim 1K $ is reached in both samples and the slope is negative beyond this point. This is similar to  ac susceptibility of spin glasses  or rather a pseudo spin glass in our case where one may identify a characteristic relaxation time $\omega_{0} \tau \sim 1$ at this temperature where $\omega_{0}$ is the device  resonant frequency. In the absence of electrons in dielectric glasses this may be interpreted as a crossover from a domination of resonant TLS interactions in the low temperature $\omega_{0}\tau\geq 1$ to a relaxation dominant regime $\omega_{0}\tau\leq 1$. In dielectric glasses the slopes on either side are expected to have a ratio $1.5$ whereas this is not valid in systems like metallic glasses\cite{akr}. 
In well studied metallic glasses \cite{akr} like PdSiCu  there is relaxation due to phonons and possibly conduction electrons as well.  The cross-over temperature is much higher $\sim 1.5\textrm{ }K$ which is large compared to dielectric glasses where it is $\sim 0.2\textrm{ }K $. In ref\cite{akr} the $T_{co}~1.5\textrm{ }K  $ is not a simple cross over from resonant to relaxation  as it does not scale significantly with frequency for a device of $1000$ Hz and $1$ GHz whereas dielectric glasses do satisfy a frequency scaling, $\omega T_{co}^3$ being a constant. We do not have several decades of scaling as in \cite{akr}, but we do not see a $\Delta T_{co} \sim 75\textrm{ }mK $ between a $19$ MHz and $29$ MHz beams. A similar cross over $\sim 1K$ was seen for Al resonators from $40 \textrm{ to } 350$ MHz \cite{hoehne}. However the slope on the right hand side of the crossover is not linear as seen in metallic glasses\cite{akr}. The ratio of slopes at the cross over is $\sim 2$ and $2.5$ as opposed to $1.5$ for di-electric glasses. 
 In poly-crystalline metals generally and also $Pd$ films one finds $T_{co}\sim 60\textrm{ }mK$ as reported in \cite{esqui_metals}. The same work reported no dependence of the TLS phonon coupling constant $\gamma$ on electronic mean free path and $\gamma$ was generally $1\textrm{ to }3\textrm{ }eV$ for metals. An increase of $T_{co}$ by $50\textrm{ } mK $ for annealed Pt films was interpreted as a decrease in the TLS-phonon coupling constant $\gamma.$ Annealed films have tensile stress. Since our beams are under tensile stress, we may conclude that $\gamma$ is reduced in comparison to bulk films.The dissipation shows a sub-linear power law  in both cases below the $T_{co}$. The power law shows a weak change above $T_{co}$. The deviation from the power law and the onset of a characteristic saturation is obvious at higher temperatures.  Overall the behaviour shows features of glass like TLS models but does not fit either a dielectric glass or metallic glass.
 
The scenario with low $H_2$ concentration Fig(\ref{b}) and  high $H_2$ concentration Fig(\ref{c})  is qualitatively similar with some key differences. In cases of  both low and high $H_2$ concentrations  the samples showed a clear lowering of the $T_{co}$ to around $700\textrm{ } mK $. It is important that for devices close to  $\sim 20$ MHz in both of these samples we see a lower cross over temperature as compared to the $H_2$ free case. 

 The dissipation shows a sub-linear power-law around $0.37-0.43$ . It is easy to adjust the power law by choosing or neglecting a few points near the $T_{co}$ in frequency due to scattering from onset of a weaker power or plateau like feature near $T_{co}$. But this dependence is not systematic with or without $H_2$, hence we quote a mean of $\sim T^{0.4}$ to describe the behavior. There were no Debye peaks\cite{krim} in the dissipation indicating that $H_2$ diffusion induced mechanisms are not prominent at these temperatures which are also well below freezing point of $H_2$. The overall order of magnitude of dissipation is not significantly lower despite of the softening seen by lower power for probing in the linear response regime. If we take the dissipation in the high temperature plateau or the highest temperature, where the data show a change to weaker dependence the constant C from  TLS theory turns out to be similar order 
\begin{displaymath}
C = \left\{ \begin{array}{ll}
\textrm{if } T > T_{co} &
\left(\frac{2}{\pi}\right)Q^{-1} \sim \frac{P\gamma^2} {E}\sim 0.1 \textrm{ to } 5 \times 10^{-5}   \\
\textrm{ if } T< T_{co} & \left(\frac{df}{f_{0}}\right) \sim 1.5 \textrm{ to } 4\times 10^{-5} \\
\end{array}\right.
\end{displaymath} \label{tls_eqn}
in the $H_2$ free and the low $H_2$ scenario. This similarity is surprising considering the effective Young's modulus $E$ has to be lower in the case of the $H_2$ scenario as reckoned by lower drive power to keep the response linear. The only plausible explanation is either the TLS density $P$ or phonon coupling $\gamma$ is enhanced. The standard TLS models reckon $\omega \sim \gamma^2 T_{co}^3$. The lowering of $T_{co}$ in our case implies enhancement of $\gamma$ due to additional compressive stress caused by $H_2$ adsorption. This is reminiscent of materials like PMMA (softer glass) and silica (harder glass)  showing similar low temperature dissipation due to commensurate TLS parameters for $C$\cite{tls_stress}. Surprisingly the ratio of frequency shifts above and below $T_{co}$ is $ \sim1.43$ and $\sim1.5$ for the low $H_2$ samples in Fig(\ref{b}b) similar to amorphous dielectrics. 

 In  the low $H_2$ scenario Fig(\ref{b}a) at temperatures below $200\textrm{ } mK $ one sees a saturation in dissipation. Such a feature has  been predicted \cite{ahn} as a possible super-radiant  phonon emission. In some older experiments\cite{gold} where the temperature was corrected using  the high temperature logarithmic frequency shift as a thermometer the feature merged into the power law indicative of thermal de-coupling and these features were at much lower temperatures \cite{gold,hakonen}.  Such a correction also did not remove the feature as seen in Fig(\ref{b}.a). The sample with higher $H_2$ Fig(\ref{c}a) shows the same feature starting earlier at $\sim300\textrm{ } mK $ down to $\sim 250\textrm{ }mK$ where we could get the data without any need for corrections to the frequency shift.
 In ref\cite{ahn} two characteristic loss mechanisms are added $Q^{-1}_{total}= Q_{0}^{-1}  + Q^{-1}_{pump}$ where  $Q_{0}^{-1}$ is the  TLS mechanism and $Q^{-1}_{pump}$ a co-operative phonon emission by a small population of  TLS enclosed within phonon wavelength. The effect is a weak renormalization of TLS relaxation time $\tau$ by $\tau/N$ where $N$ is the number of  TLS that are cooperatively excited by a phonon.   
 The effect is expected only at very low temperatures due to crossover to non-linear TLS-phonon coupling. The high temperature plateau are different where $\omega\tau\sim 1 $ signifies onset of inelastic processes exciting a spectrum of TLS relaxation rates. When already in the limit $\omega \tau > 1$ we expect coherent excitation. One of the criterion for the pumping\cite{ahn} is the asymmetry energy $\Delta_{0} \leq \gamma \epsilon_{o}$ where $\gamma $ is the TLS phonon coupling  and $\epsilon_{o}$ some in built strain. We cannot comment on the static strain status of our devices at cryogenic temperatures but devices exposed to $H_2$ showed some buckling when warmed up and the power to drive them linearly was significantly lower. We also inferred $\gamma$ to be enhanced from drop in $T_{co}$ with $H_2$ exposure. The condition to probe more TLS by a phonon is also enhanced on a lattice with compressive strain. Other possible scenarios like modification of  TLS distribution or interactions  are expected at ultra low temperatures, typically below $10 \textrm{ }mK$\cite{feffer}. In our data the small frequency shift corrections in Fig(\ref{b}a.) or none in Fig(\ref{c}) for higher $H_2$ sample points towards a super-radiant phonon loss as the most plausible scenario. 

In conclusion we have studied a system where dissipation scenarios are tunable. There is prospect for further studies at KHz frequencies to several  hundred MHz  and other experimental probes like thermal transport. Further experiments in these extreme regimes have the potential to throw light on TLS like mechanisms in sub-micron metallic structures.    
\begin{acknowledgments}
 AV thanks DST Nanomission, DST Ramanujan Fellowship and IISER Mohali for funds. We also thank  DST Inspire Fellowships and CSIR-UGC for funding  students. We would like to thank Mrs C Linz and Dr J. Eroms for help with HF etching and Mr Inderjit Singh for help in clean-room. AV thanks Prof A.K. Raychaudhuri for several insightful discussions. We acknowledge critical review of the manuscript by  Prof J.R Owers-Bradley Prof A.D. Armour, Dr A. Huettel and Dr P. Balanarayan. We thank Dr S. Goyal for some discussions on super-radiance.
\end{acknowledgments}


\begin{thebibliography}{99}

\bibitem{schwab} K.C. Schwab and M.L. Roukes Physics Today {\bf 58} , 36 (2005). 

\bibitem{weig} E. M. Weig, R. H. Blick, T. Brandes, J. Kirschbaum, W. Wegscheider, M. Bichler, and J. P. Kotthaus Phys. Rev. Lett. {\bf 92}, 046804 (2004).

\bibitem{spin_flip}G. Zolfagharkhani, A. Gaidarzhy, P. Degiovanni,S. Kettemann, P. Fulde and P. Mohanty Nature\ Nanotechnology\ {\bf 3}, 720 - 723 (2008). A. G. Mal’shukov, C. S. Tang, C. S. Chu, and K. A. Chao Phys. Rev. Lett. {\bf 95}, 107203 (2005).

\bibitem{ekinci} K. L. Ekinci and M. Roukes, Rev. Sci. Instrum. {\bf 76}, 061101 (2005).

\bibitem{huang}  X. M. H. Huang, X. L. Feng, C. A. Zorman, M. Mehregany, and M. L. Roukes, New Journal of Physics {\bf 7}, 247 (2005) ; 
 X.M.H. Huang, C.A. Zorman, M. Mehregany and M.L. Roukes   Nature {\bf 421}, 496 (2003).

\bibitem{gaid} A. Gaidarzhy, G. Zolfagharkhani, R. L. Badzey, and P. Mohanty, Phys. Rev. Lett. {\bf 94}, 030402 (2005).


\bibitem{lehnert} J. D. Teufel, J. W. Harlow, C. A. Regal, and K. W. Lehnert
Phys.\ Rev.\ Lett.\ {\bf 101}, 197203 (2008). T. Rocheleau, T. Ndukum,C. Macklin, J. B. Hertzberg, A. A. Clerk  \& K. C. Schwab Nature {\bf 463}, 72-75 (2010).
\bibitem{connel} A. D. O'Connell, M. Hofheinz, M. Ansmann, Radoslaw C. Bialczak, M. Lenander, Erik Lucero, M. Neeley, D. Sank, H. Wang, M. Weides, J. Wenner, John M. Martinis and A. N. Cleland Nature {\bf 464}, 697-703 (2010). 
\bibitem{remus} L. G. Remus, M. P. Blencowe, and Y. Tanaka, Phys.\ Rev.\ B\ {\bf 80},
174103 (2009).

\bibitem{enss} C.H. Enss and R. Hunklinger \emph {Low Temperature Physics} Springer-Verlag, Berlin-Heidelberg (2005). 

\bibitem{feffer} A. D. Fefferman, R. O. Pohl, A. T. Zehnder, and J. M. Parpia Phys.\ Rev.\ Lett.\ {\bf 100}, 195501 (2008).


\bibitem{esqui} P. Esquinazi ed. \emph{Tunneling Systems in Amorphous and Crystalline Solids}  Springer-Verlag, Berlin (1998).

\bibitem{hunk_akr}  S. Hunklinger and A. K. Raychaudhuri, {\emph Progress in Low Temperature Physics},
ed. by H. Brewer Elsevier, New York,Vol IX (1986).

\bibitem{liuh2} Xiao Liu, B. E. White, Jr., R. O. Pohl, E. Iwanizcko, K. M. Jones, A. H. Mahan, B. N. Nelson, R. S. Crandall, and S. Veprek Phys.\ Rev.\ Lett.\  {\bf 78}, 4418 (1997).

\bibitem{hellman} X. Liu, D.R. Queen, T,H. Metcalf, J.E. Karel, and F. Hellman
Phys. Rev. Lett. {\bf113}, 025503 (2014).

\bibitem{anderson_varma} P.W. Anderson, , B.I. Halperin,C.M. Varma, Philos.\ Mag.\ {\bf 25},
1 (1972).

\bibitem{parpia1} D. R. Southworth, R. A. Barton, S. S. Verbridge, B. Ilic, A. D. Fefferman, H. G. Craighead, and J. M. Parpia Phys.\ Rev.\ Lett.\ {\bf 102}, 225503 (2009).

\bibitem{mohanty_rev} M. Imboden \& P. Mohanty  Physics\ Reports\  {\bf 534}  89–146 (2014).


\bibitem{leggett} A.J. Leggett and D.C. Vural | J. Phys. Chem. B , {\bf 117}, 12966 (2013).

\bibitem{yu} C. C. Yu, J. Low Temp. Phys. {\bf 137}, 251 (2004).

\bibitem{lulla} K. J. Lulla, M. Defoort, C. Blanc, O. Bourgeois, and E. Collin
Phys.\ Rev.\ Lett.\ {\bf 110}, 177206 (2013).

\bibitem{li} T. F. Li, Yu. A. Pashkin, O. Astafiev, Y. Nakamura, J. S. Tsai, and
H. Im, Appl.\ Phys.\ Lett.\ {\bf 92}, 043112 (2008).



\bibitem{gold} A. Venkatesan, K. J. Lulla, M. J. Patton, A. D. Armour, C. J. Mellor, and J. R. Owers-Bradley Phys.\ Rev.\ B. {\bf 81}, 073410 (2010);A. Venkatesan, K. J. Lulla, M. J. Patton, A. D. Armour, C. J.Mellor, and J. R. Owers-Bradley, J.\ Low Temp.\ Phys.\ {\bf 158}, 685
(2010).



\bibitem{hoehne} F. Hoehne, Yu. A. Pashkin, O. Astafiev, L. Faoro, L. B. Ioffe, Y. Nakamura, and J. S. Tsai Phys.\ Rev.\ B {\bf 81}, 184112 (2010).

\bibitem{hakonen} J. Sulkko, M. A. Sillanpää, P. Häkkinen, L. Lechner, M. Helle, A. Fefferman, J. Parpia, and P. J. Hakonen Nano Lett.{\bf 10} (12), pp 4884–4889 (2010).

\bibitem{tls_stress} J. Wu and C.C. Yu Phys.\ Rev.\ B.\ {\bf 84}, 174109 (2011)

\bibitem{h2} X. M. H. Huang, M. Manolidis, Seong Chan Jun1 and J. Hone Appl.\ Phys.\ Lett{\bf 86}, 143104 (2005).



\bibitem{cleland} A. Cleland and M.L. Roukes Sensors and Actuators {\bf72} 256–261 (1999).

\bibitem{atalaya} J. Atalaya, A. Isacsson, and M. I. Dykman Phys.\ Rev.\ Lett.\ {\bf 106}, 227202 (2011).
\bibitem{krim} A. Dayo, W. Alnasrallah, and J. Krim Phys.\ Rev.\ Lett.\ {\bf 80}, 1690 (1998). 


\bibitem{akr} A.K. Raychaudhuri and S. Hunklinger Z.\ Phys.\ B \- Condensed Matter {\bf 57}, 113-125 (1984).

\bibitem{esqui_metals} E. Gaganidze and P. Esquinazi  J.\ Phys.\ IV France\ {\bf 06}  C8 515 (1996).

\bibitem{ahn} K.H. Ahn and P. Mohanty  Phys.\ Rev.\ Lett.\ {\bf 90}, 085504 (2003) . 







\end{thebibliography}
\end{document}